\documentclass[11pt,english]{article}          
\usepackage{babel}
\usepackage[T1]{fontenc}

\usepackage{graphicx}
\usepackage{mathptmx}      
\usepackage{flushend}
\usepackage{wasysym}
\usepackage[numbers,sort&compress]{natbib}
\usepackage[colorlinks,citecolor=blue,urlcolor=blue,linkcolor=blue]{hyperref}
\usepackage{geometry}
\usepackage{xspace,upgreek}

\usepackage{amsmath}
\DeclareMathOperator\arctanh{arctanh}

\newcommand{\pT}{\ensuremath{p_\mathrm{T}}\xspace}
\newcommand{\pTm}{\ensuremath{p_\mathrm{T}^{\mu}}\xspace}
\newcommand{\pTmm}{\ensuremath{p_\mathrm{T}^{\mu\mu}}\xspace}
\newcommand{\pTp}{\ensuremath{p_\mathrm{T}^p}\xspace}
\newcommand{\Mmm}{\ensuremath{M_{\mu\mu}}\xspace}

\begin{document}

\title{Exclusive Lepton Pair Production at the Electron-Ion Collider -- A Powerful Research Tool }

\author{Janusz J. Chwastowski$^a$, 
Krzysztof Piotrzkowski$^b$ \\ and Mariusz Przybycien$^b$
\footnote{Corresponding authors:\newline Janusz.Chwastowski@ifj.edu.pl,~Krzysztof.Piotrzkowski@agh.edu.pl,~Mariusz.Przybycien@agh.edu.pl}\\[8pt]
$^a$ The Henryk Niewodnicza\'nski Institute of Nuclear Physics,\\
Radzikowskiego 152,
31-342 Krak\'ow, Poland\\[4pt]
$^b$ AGH University of Science and Technology, \\
Faculty of Physics and Applied Computer Science, \\
al. A. Mickiewicza 30, 30-059 Krak\'ow, Poland
}

\maketitle

\abstract{The two-photon exclusive production of lepton pairs at the Electron-Ion Collider will 
open interesting research directions thanks to a very high luminosity and clean experimental 
conditions.  A survey of the scientific potential of such studies is reported. In particular, 
we consider unique measurements of the proton elastic electromagnetic form-factors and 
a possibility of studying the anomalous electromagnetic dipole moments of $\tau$ leptons.}
\ \\[4pt]

 
\section{Introduction}\label{sec1}
The future Electron-Ion Collider and its experiment(s) 
will provide perfect conditions for studying exclusive processes:
(a) a very high luminosity will ensure high statistics data even for relatively rare processes, 
(b) the data streaming will result in no trigger losses and in a lack of the efficiency corrections, 
(c) negligible event pileup, excellent particle momentum resolutions and the particle identification (at low and medium transverse momenta) will strongly enhance full final state reconstruction \cite{YR}. 
In addition, in the far-forward and far-backward directions, high resolution detectors of protons and electrons, respectively, will enable the ``over-constrained kinematic event reconstruction'', resulting also in possibility of precise data-driven inter-calibrations and tests of the understanding of acceptances and reconstructions.

Thanks to a very high $ep$ luminosity at the EIC, very large samples of exclusively produced lepton pairs can be acquired, opening important research directions -- in the following we will mostly discuss the exclusive two-photon production of muon pairs. These first exploratory studies are performed using true kinematic variables, neglecting the detector effects, except from their expected geometrical and kinematic acceptances \cite{YR}.
\section{Monte Carlo event generation and detector acceptance}\label{sec2}

{\sc Grape} \cite{grape} is a Monte Carlo event generator for lepton pair production in electron-proton collisions. The considered processes of di-lepton  creation are $\gamma\gamma$, $\gamma Z$, $ZZ$ and internal photon conversions. Also, effects of the on-/off-shell $Z$ boson production are included, as well as those of the Initial and Final State Radiation (ISR/FSR). The cross-section is calculated using the exact matrix elements at the tree level.  The non-zero fermion masses are used in the matrix elements and kinematics. It is also possible to select a sub-set of the diagrams in the calculations.
The generator includes also the $e^\pm e^\pm$ interference in the $e^+e^-$ channel. The proton kinematics considers three classes of the scattering 
processes: elastic, quasi-elastic and DIS. 

Below only the elastic case is studied where the proton-proton-photon vertex is calculated using the standard formulae exploiting dipole representation of the proton electromagnetic form-factors:
\begin{equation}
G_E(t)=(1-t/A_0)^{-2},\;\;\;\;\;\; G_M(t)=\mu_pG_E(t)
\label{eq:ff}
\end{equation}
where $t$ denotes the four-momentum transfer squared at the proton vertex, $A_0=0.71~{\rm GeV}^2$ and $\mu_p$ is the proton magnetic moment.

The effect of ISR is included in the cross-section calculation using the structure 
function method described in \cite{grape-isr}. FSR is performed by 
{\sc Pythia~6.4}~ \cite{pythia64} using the parton shower method.\\~

\noindent
The generated events were required to fulfill the following selection:
\begin{equation}
\label{eq:cuts}
    \begin{array}{ll}
      1)  &0.5 < E^\prime_e/E_e < 0.9 \textrm{ and } \pi-\theta_e < 10 \textrm{ mrad for the scattered electron,}  \\
      2)  &x_L < 0.97 \textrm{ or } \pTp > 100 \textrm{ MeV}/c, \textrm{ and } \theta_p < 13 \textrm{ mrad for the scattered proton,} \\
      3)  &\pT^\ell > 300 \textrm{ MeV/}c \textrm{ and }\lvert \eta_\ell \rvert < 3.5 \textrm{ for leptons belonging to a pair,}\\
      4) &\textrm{photon veto: no photons above 200 MeV within } \lvert\eta\rvert<4,
    \end{array}
\end{equation}
with $x_L = p^p_z/P_p$, where $P_p$ is the proton beam momentum.
The first three demands introduce the EIC detector acceptance and constitute a set of standard requirements. The fourth one, removing events with hard FSR, is sometimes imposed to improve the event kinematics reconstruction.

\section{Results}\label{sec3}
\begin{figure}[h]
\centering
\includegraphics[width=0.49\textwidth]{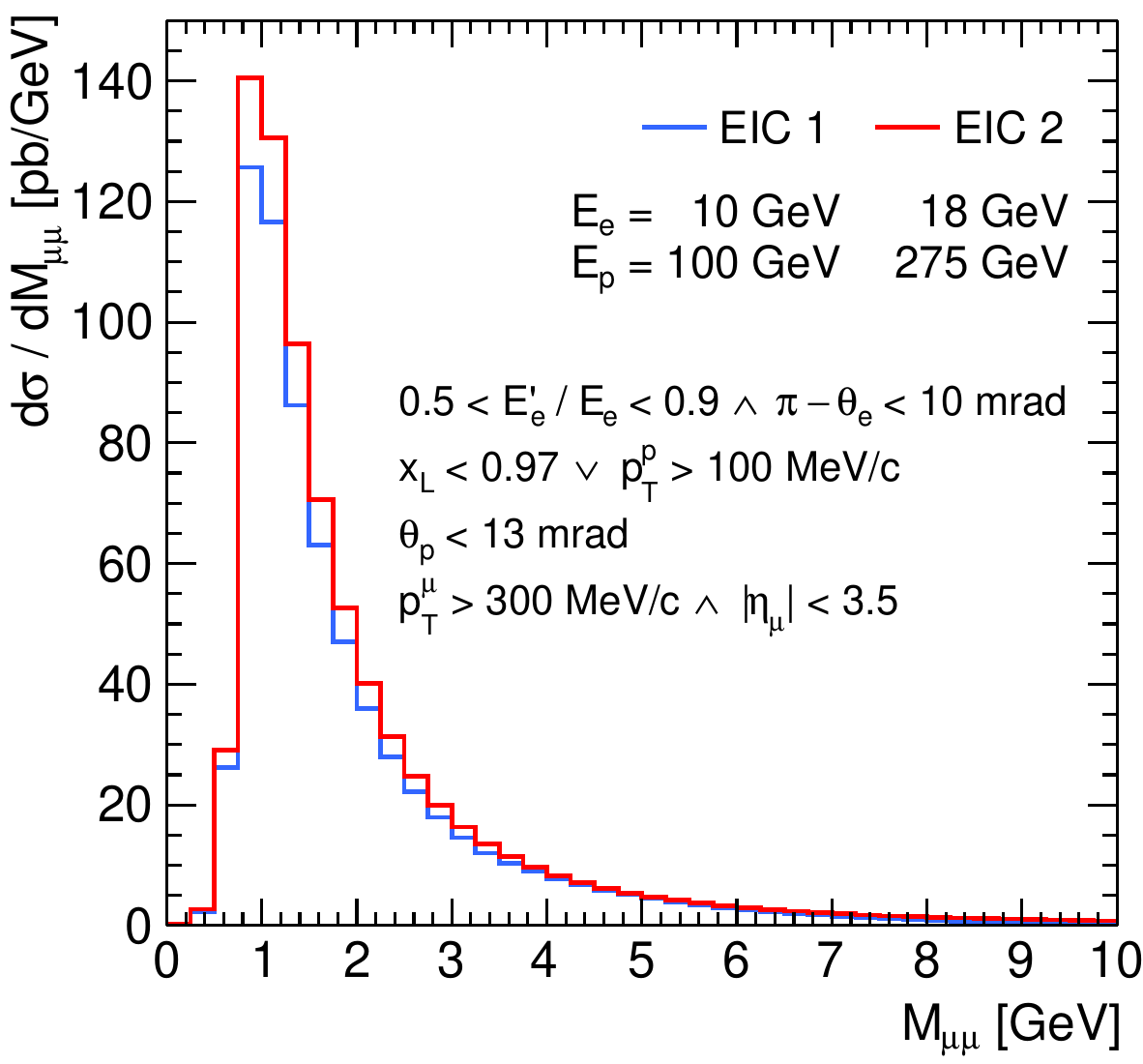}
\includegraphics[width=0.49\textwidth]{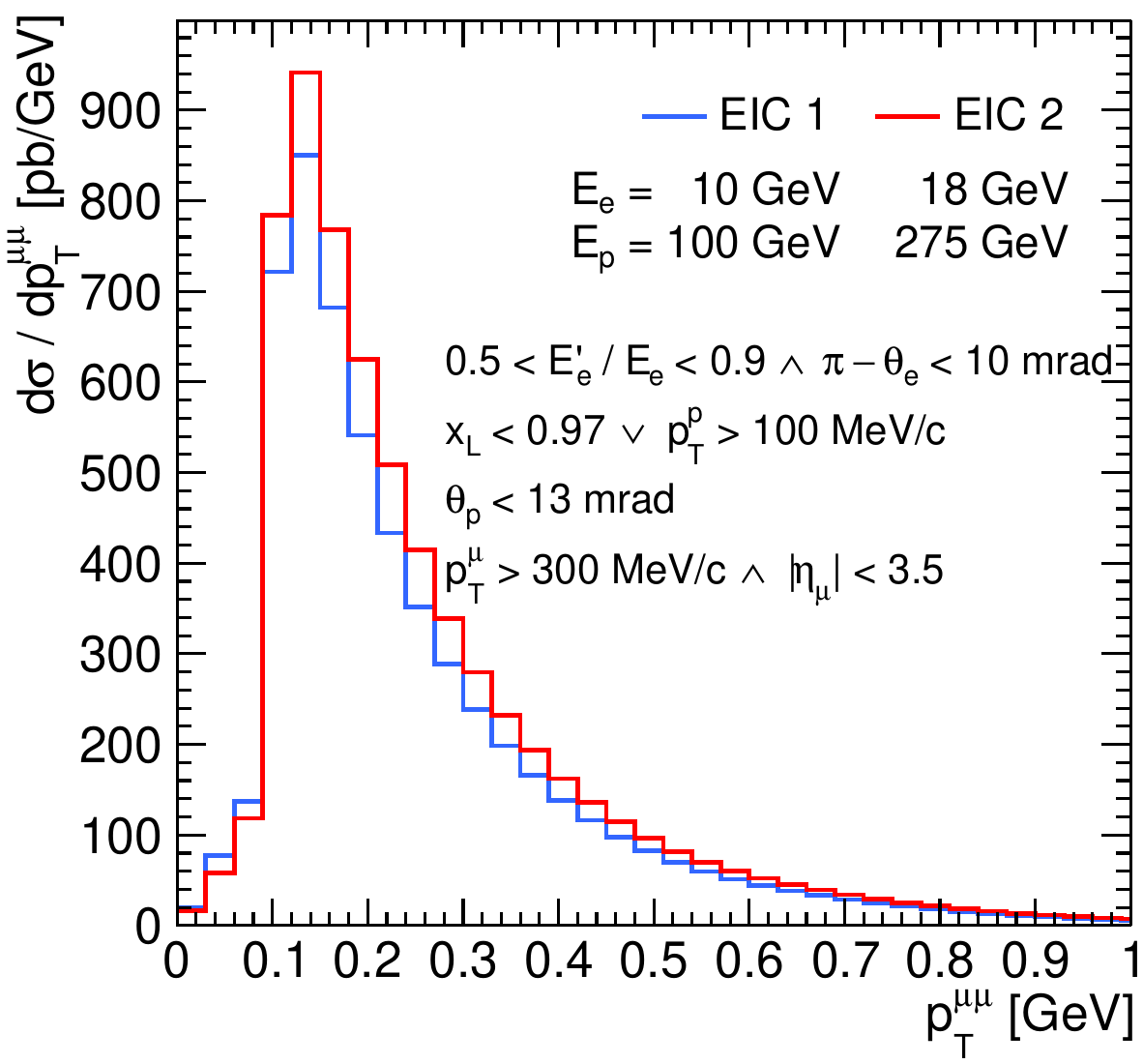}
\caption{(a) The ``observed" cross-section as a function of the invariant mass \Mmm for the selected events according to the set of cuts \ref{eq:cuts}, without the photon veto, for the high (red histogram) and low (blue histogram) energy collisions at the EIC; (b) the corresponding cross-sections for the muon pair as a function of \pTmm.}  \label{fig1}
\end{figure}

In Fig. \ref{fig1} the differential cross-sections are shown for the muon pairs passing the selections cuts without the photon veto, at the two energy settings of the EIC:
EIC~1 (electron beam energy $E_e=10$ GeV and proton beam energy $E_p=100$~GeV) and EIC~2 ($E_e= 18$ GeV and $E_p=275$~GeV). Visible threshold effects at the low invariant mass \Mmm and transverse momentum of a pair, \pTmm, are due to requirements on the minimal muon \pTm and the acceptance of the scattered proton, respectively. The observed total cross-sections are equal to 169~pb and 192~pb (163~pb and 185~pb if the photon veto is applied) for low and high energy beams, respectively. This shows that, given very high luminosities expected at the EIC, the available statistics of exclusive lepton pairs will be very large, even when the full final state is detected. 

One should note that clean samples of exclusive muon pairs were collected not only at HERA but also in very harsh conditions at the LHC. The cosmic ray muons can be very efficiently suppressed by tight vertex and acolinearity cuts. In addition, the planned technique of data streaming at the EIC will remove sensitivities to the trigger algorithms as well as maximize the muon acceptance.
\subsection{Calibration of far-forward and far-backward detectors} \label{sec3.1}
\begin{figure}[h]%
\centering
\includegraphics[width=0.49\textwidth]{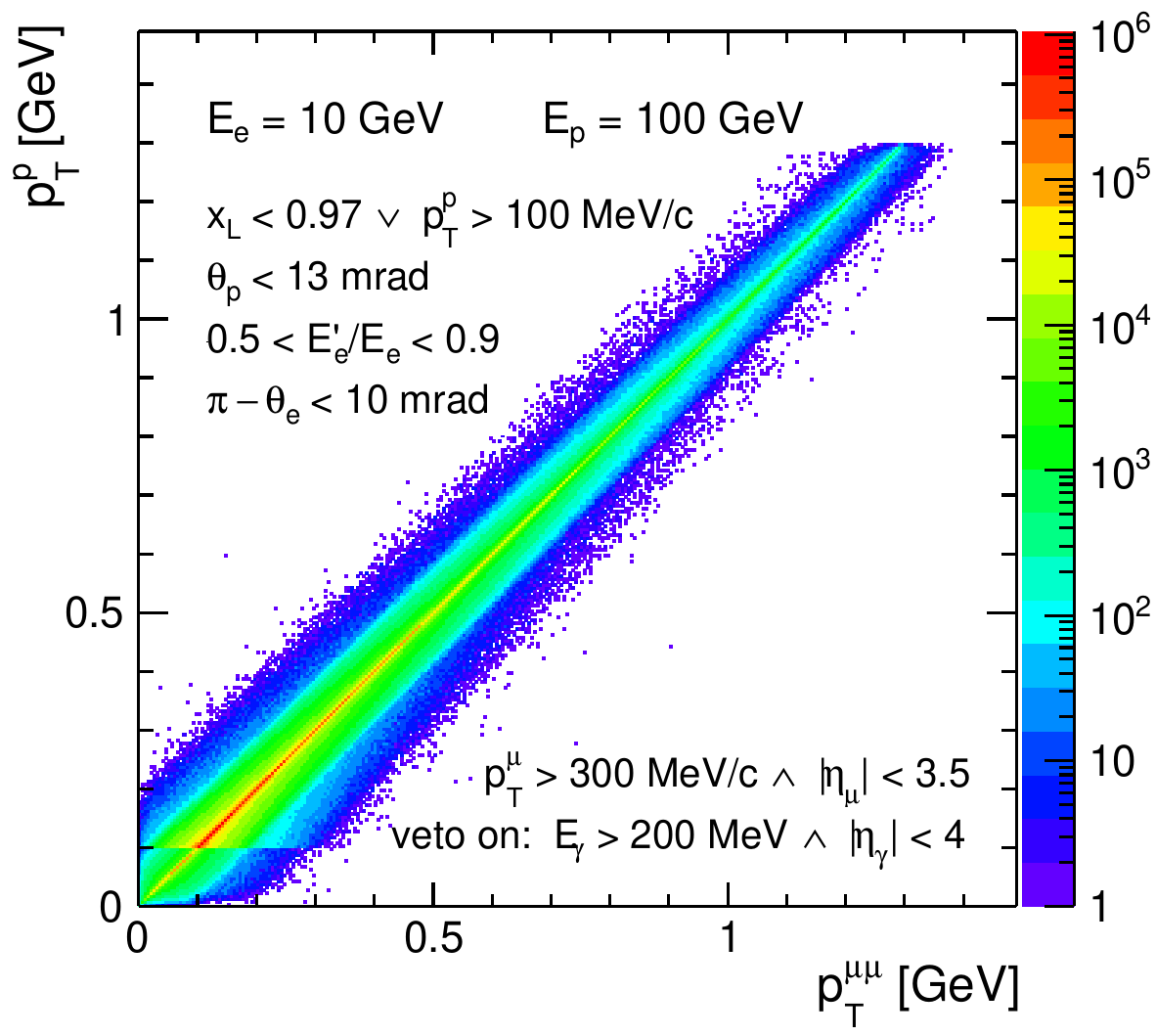}
\includegraphics[width=0.49\textwidth]{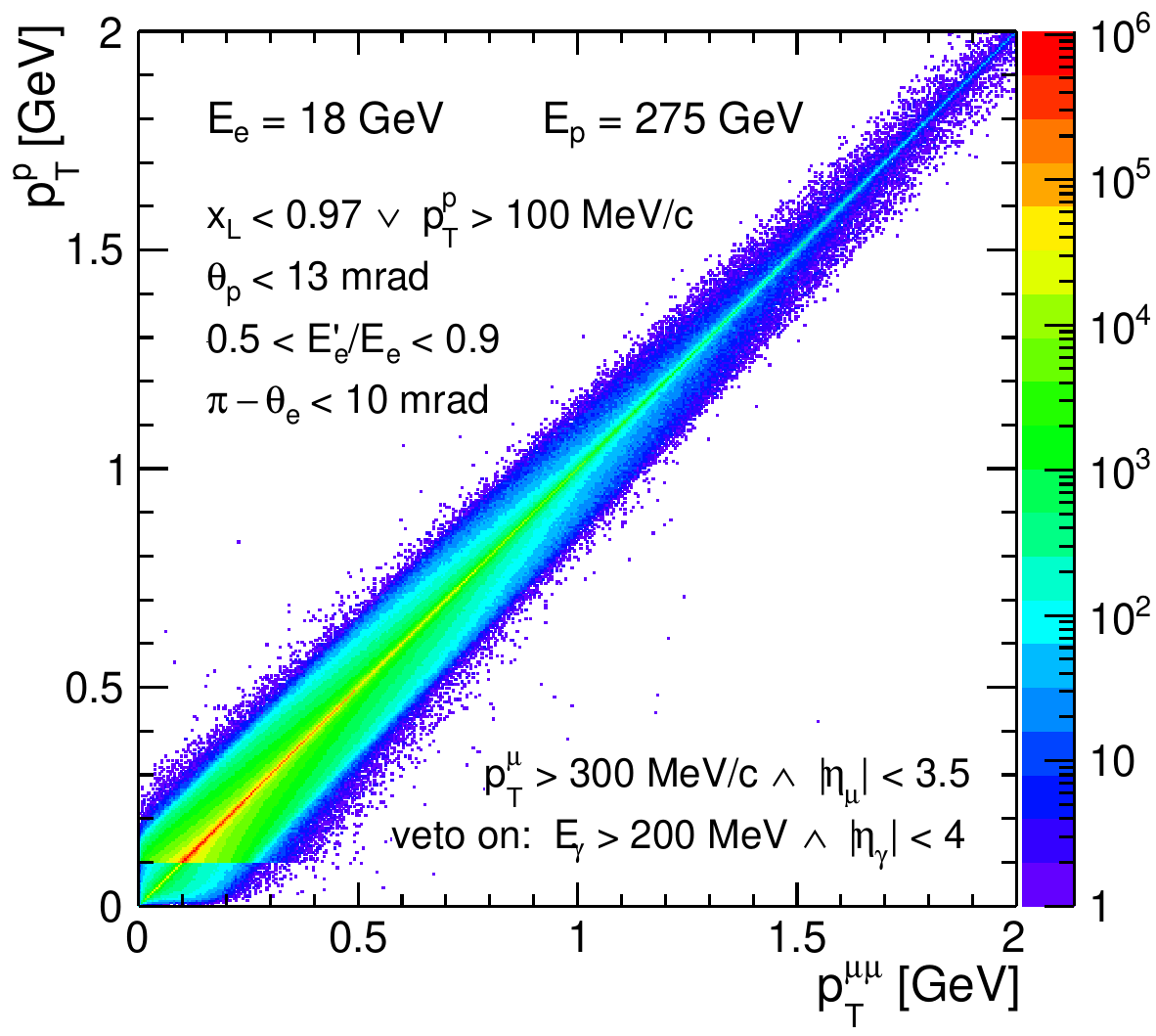}
\caption{Correlation of the lepton pair transverse momentum, \pTmm, and the proton transverse momentum, \pTp for two energy configurations of the EIC. The photon veto (\ref{eq:cuts}.4) was applied. 
} 
\label{fig2}
\end{figure}

The exclusive muon (and electron) pairs can be used as a powerful tool for calibrating the far-forward and far-backward detectors at the EIC. In Figure~\ref{fig2} the two-dimensional distributions of the muon pair transverse momentum \pTmm and the proton transverse momentum \pTp demonstrate a very strong linear correlation, amplified by the requirement of the scattered electron within acceptance of the far-backward detectors, as that ensures a very small \pT transfer at the electron vertex (or, equivalently, a very low photon virtuality $Q^2$). 
The longitudinal momenta of the scattered electrons and protons can be calibrated following the Drell-Yan technique used for determination of the fractional momenta of collinear partons from the invariant mass and rapidity of lepton pairs: 
$$x_{1,2} = \frac{M_{\ell\ell}}{\sqrt s}\sqrt{\frac{(E^{\ell\ell}\pm p^{\ell\ell}_z)}{(E^{\ell\ell} \mp p^{\ell\ell}_z)}}\exp{(\mp Y^*)},$$
where $x_1, x_2$ are the fractional longitudinal momenta transfers at the electron and proton vertices, respectively, $E^{\ell\ell}$ and $p^{\ell\ell}_z$ are the lepton pair energy and longitudinal momentum, and 
$$
Y^* = \arctanh{\left(\frac{P_{e,z}+P_{p,z}}{E_{e}+E_p}\right)}
$$ 
is the rapidity of the centre-of-mass reference frame.

\begin{figure}[h]%
\centering
\includegraphics[width=0.49\textwidth]{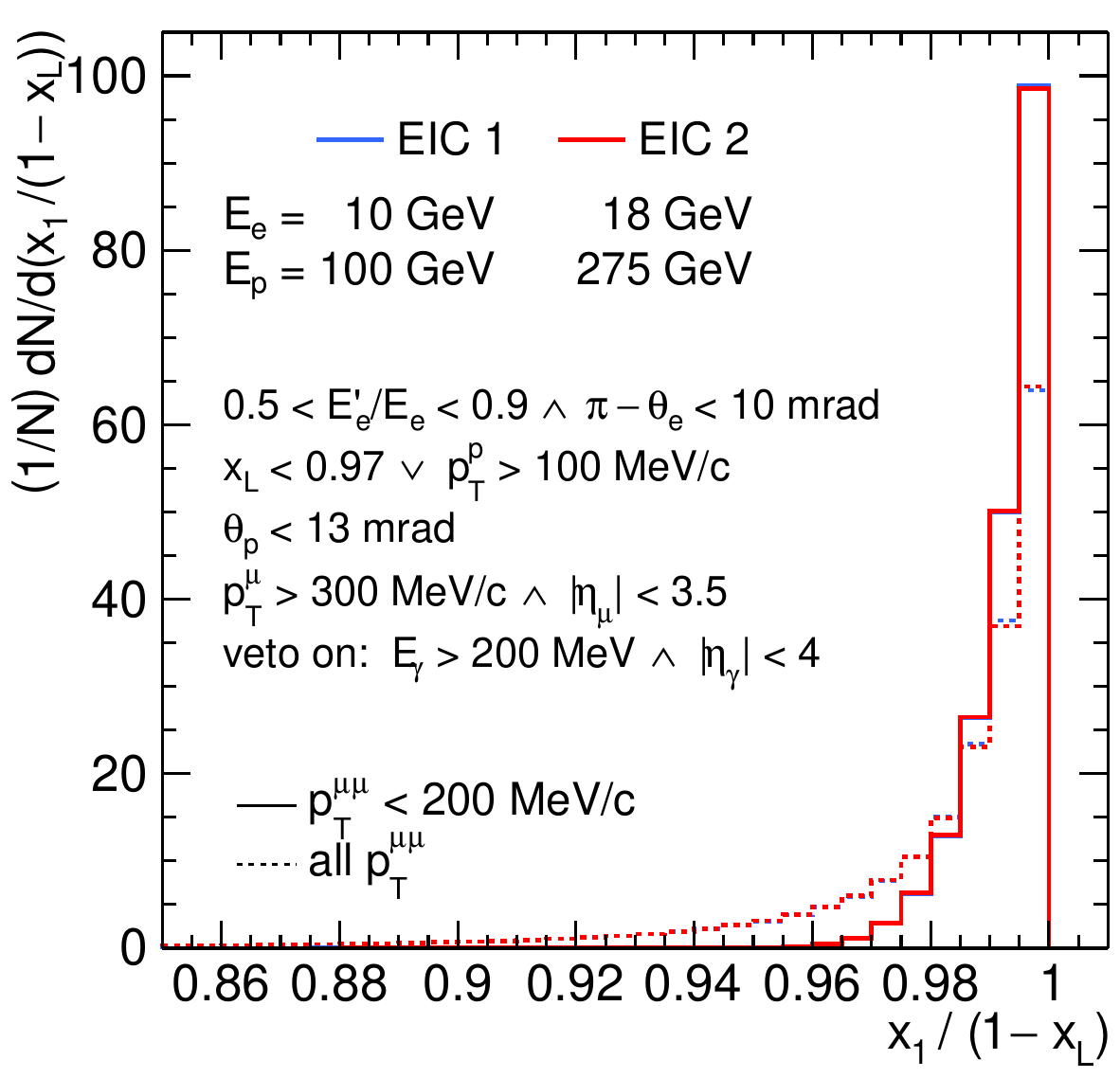}
\includegraphics[width=0.49\textwidth]{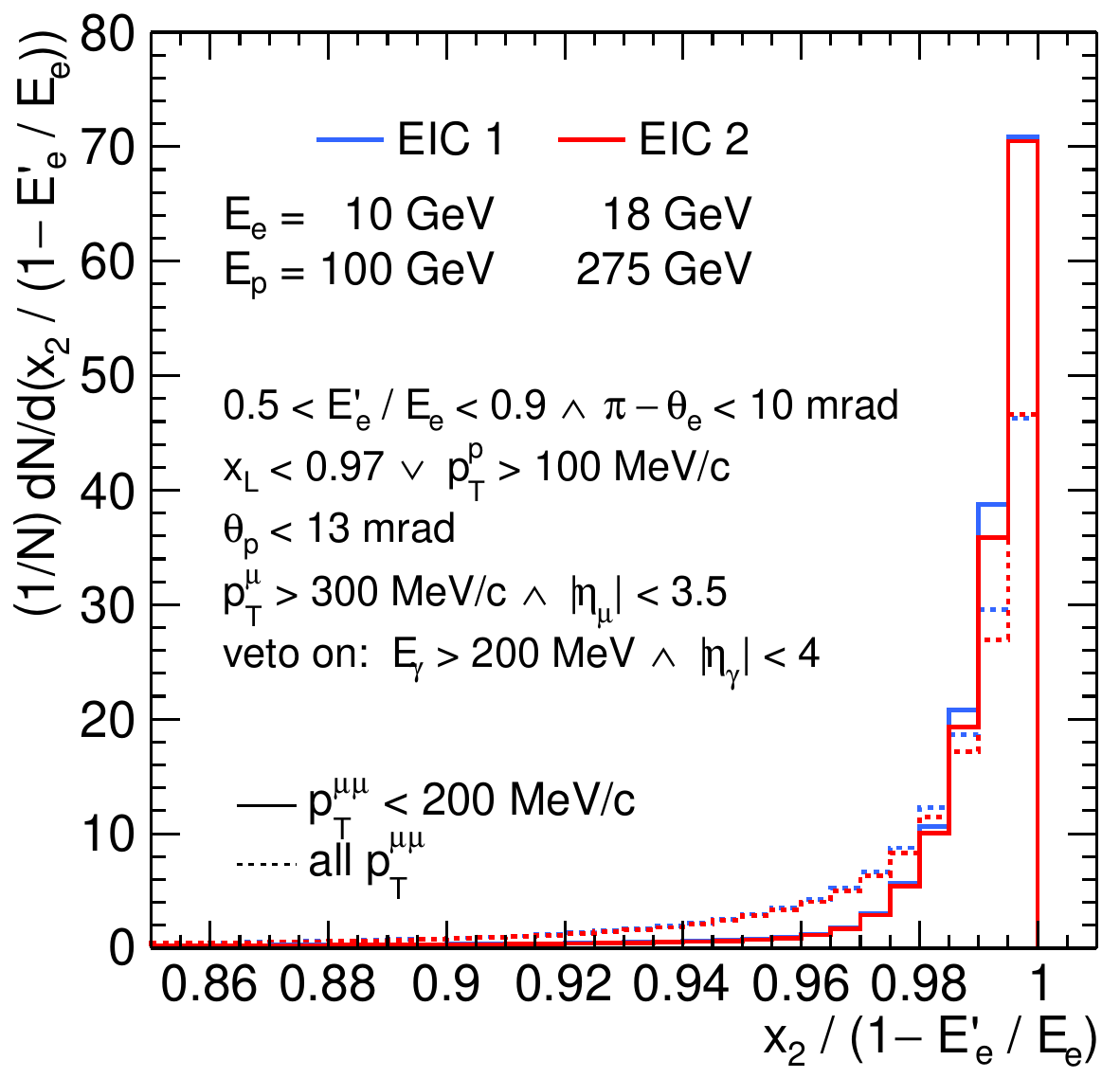}\\
\caption{Distributions of $x_1/(1-x_{L})$ and $x_2 / (1-E^{\prime}_e/E_e)$ with (solid line) and without (dotted line) the cut on the transverse momentum of the lepton pair $\pTmm<200$ MeV/$c$.} 
\label{fig3}
\end{figure}

In Fig. \ref{fig3} the ``reconstructed" $x_1, x_2$ are compared to  $1-x_L$ 
and  $ 1-E_e'/E_e$. The photon veto was applied to minimize the FSR impact, as well as the cut on the muon pair, $\pTmm<200$~MeV/c, to maximize collinearity of the exchanged photons. The obtained distributions are narrow with the FWHM below 1\%, and a sharp kinematic cut-off.

Given a large statistics of the selected sample and a high resolution of reconstruction of the muons by the central detectors, this will result in a possibility of regular precise calibrations of the far-forward and far-backward detectors. In addition, the exclusive electron-positron pairs will nearly double the statistics of such unique calibration samples. 
Moreover, as the detection of the forward-scattered proton ensures the exclusivity of the event, one can lift the requirement of the detection of scattered electron and use such an ``untagged'' sample to determine  the acceptance of the far-backward detectors directly from the experimental data.
\subsection{Sensitivity to the proton charge radius}\label{sec3.2}
There are observed continuing discrepancies among measurements of the proton charge radius $R_p$, in particular among ``classical” measurements using electron-proton elastic scattering \cite{bernauer,nature} (see also \cite{Pacetti-radius} for discussion of the experimental results). Precise measurements of the exclusive production of lepton pairs at the EIC may help to sort out this puzzle. Below, we investigate the ultimate statistical precision of the $R_p$ determination at the EIC.

The mean proton charge radius $R_p$ can be obtained from the proton electric form-factor using the following relation:
\begin{equation}
R_p^2 = 6\bigg[\frac{1}{G_E}\frac{{\rm d}G_E}{{\rm d}t}\bigg]_{t=0},
\label{eq:r}
\end{equation}
hence, $R_p^2 = 12/0.71~{\rm GeV^2}$ for the ``standard" dipole $G_E$. 
Therefore, for the dipole form-factors, the change of the 
nominal parameter $A_0 = 0.71~{\rm GeV^2}$ (Eq.~\eqref{eq:ff}) is equivalent to the 
corresponding change of the nominal value of the proton radius $R_p=0.811$~fm. 

To evaluate the statistical sensitivity to $R_p$ two ratios of the ``observed'' differential cross-sections in $t$ were calculated using three different values of the $A$ parameter in the dipole form-factors, or equivalently for three values of $R_p$. As the elastic cross-sections for the exclusive lepton pairs calculated by {\sc Grape} are proportional to $G_E^2$, therefore, close to $t=0$, their ratios are directly sensitive to the proton charge radius:
\begin{equation}
\frac{\rm d}{\rm{d}t}\left.\left(\frac{\rm{d}\sigma}{\rm{d}t}(A_1)\Big/\frac{\rm{d}\sigma}{\rm{d}t}(A_2)\right)\right\vert_{t=0} = \frac{1}{3}(R_1^2-R_2^2).
\label{eq:ratio}
\end{equation}

\begin{figure}[h]%
\centering
\includegraphics[width=0.95\textwidth]{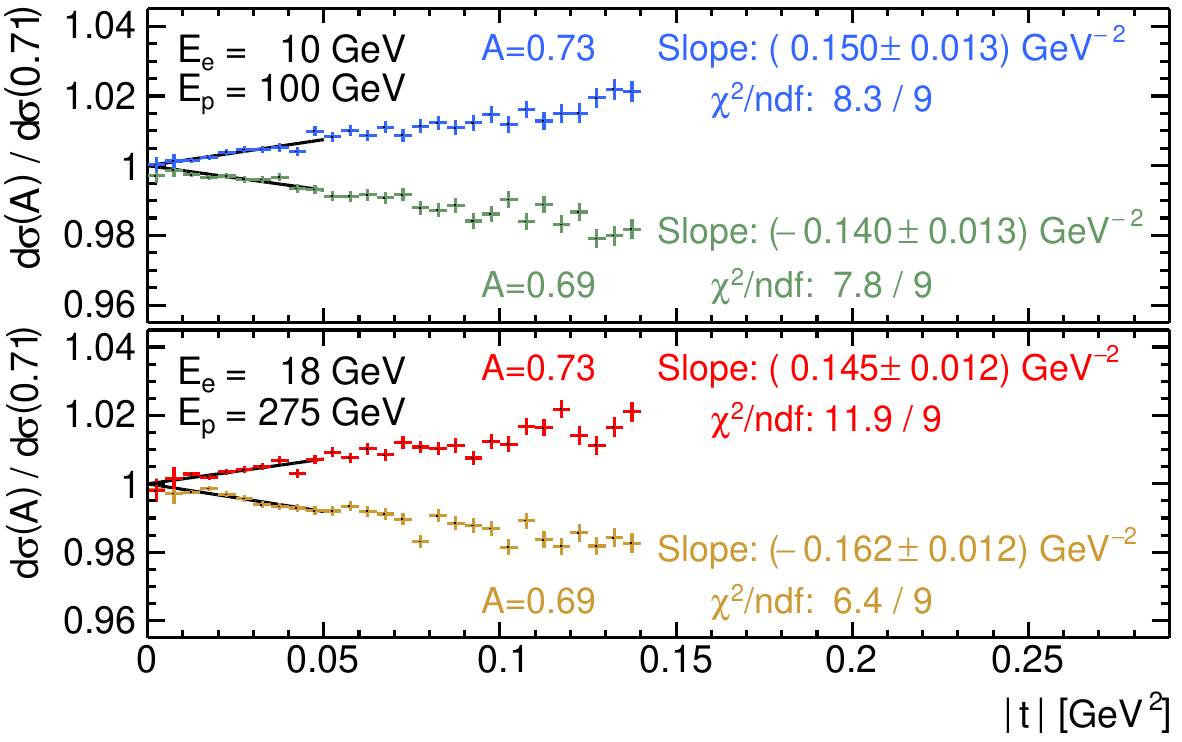}
\caption{Ratios of the observed cross-sections $\rm{d}\sigma/\rm{d}t$ for $A~=~0.69$~GeV$^2$ and 0.73~GeV$^2$ with respect to the ``standard'' $\rm{d}\sigma/\rm{d}t(A_0)$; statistical errors correspond to the integrated luminosity of 106 and 99~fb$^{-1}$ for low (upper plot) and high (lower plot) energy beams, respectively. The fitted slopes at $t \approx 0$ are also displayed. Neither electron requirement nor the photon veto were imposed (only cuts \ref{eq:cuts}.2--3 were applied).}  \label{fig4}
\end{figure}

Figure \ref{fig4} shows the ratios of the observed cross-sections -- $\rm{d}\sigma/\rm{d}t$ obtained with $A=0.69$~GeV$^2$ or 0.73 GeV$^2$, to the ``standard'' one 
for $A=A_0$. The expected derivatives at $t=0$ are equal, according to Eq. \ref{eq:ratio}, to $0.163$ and $-0.154$ for $A_1=0.69$~GeV$^2$ and $0.73$~GeV$^2$, respectively, and $A_2=A_0$. One should note that in order to increase the observed cross-sections and to avoid sensitivity to the bremsstrahlung overlays in the far-backward detectors the detection of the scattered electron was not required.

The fitted slopes in the $\lvert t\rvert$-range of $0-0.05$~GeV$^2$ agree well with the above expectations (the fitted slopes have the opposite sign to the corresponding derivatives) and the associated statistical uncertainties provide estimates of ultimate sensitivities to $R_p$. The simulated data samples correspond to the EIC integrated luminosity of about 100~fb$^{-1}$ and the estimated statistical uncertainty on $R_p$ is very small, of about 0.1\%.  

The impact of the proton beam angular divergence and and its energy dispersion was checked by "smearing" the proton momenta according to the EIC beam parameters \cite{CDR}. Fits to the smeared $t$-distributions resulted in new slopes only weakly changed, within statistical errors of the results reported in Fig. \ref{fig4}.

Resolutions of far-forward detectors will introduce additional event ``migrations'' and smearing of $t$-distributions  -- corrections for these effects, including the proton beam divergence, will result in systematic uncertainties. Such problems will be addressed in future publications; here the ultimate potential for such measurements is discussed. One should note that the total integrated luminosity of the order of 1000~fb$^{-1}$ is expected at the EIC, and in addition the exclusive electron-positron pairs can also be used for such studies. That will lead to yet much bigger data samples, taken at various beam energies and polarizations, allowing for many powerful studies of potential systematic effects.

The above demonstrates that the measurements of exclusive lepton pairs at the EIC may 
enable unique determinations of the proton charge radius with competitive precision. 
The proposed technique can also be applied to perform novel measurements of the elastic form-factors and charge radii of light ions, such as the deuterium and helium nuclei, for which the far-forward detectors at the EIC  will still have significant acceptances and for which the experimental data show some tension \cite{Pohl:2016xsr,HERNANDEZ2018377}.

In the case of heavy nuclei, as the gold nuclei, the elastically scattered ions cannot be detected, therefore the $t$-dependence of the cross-section will be measured using the transverse momentum of the lepton pair. 
In this case, the tagging of low-$Q^2$ events using the far-backward detectors might be mandatory to get the best $t$ resolution. 
The major systematic problem will appear, however, because of the contribution of non-elastic (or semi-exclusive) events, when an incident ion gets excited or dissociates into a system of a larger mass, as a result of an interaction.

\subsection{Electromagnetic form-factors at higher photon virtualities}
\label{sec3.3}

Precise measurements of the proton electromagnetic form-factors are also of interest at larger $\lvert t\rvert$, at $1~{\rm GeV}^2$ and beyond, where the perturbative calculations can be performed. These two elastic form-factors are contained in the Generalized Parton Distributions (GPDs) (see for example \cite{Belitsky:2005qn}). However, the GPDs cannot yet be determined fully from first principles the high precision data will deliver an experimental platform  for further constraining of them, cf.  \cite{Diehl:2004cx} or can be used to discuss the lattice QCD predictions, e.g.~\cite{Alexandrou:2018sjm}.

\begin{figure}[h]%
\centering
\includegraphics[width=0.7\textwidth]{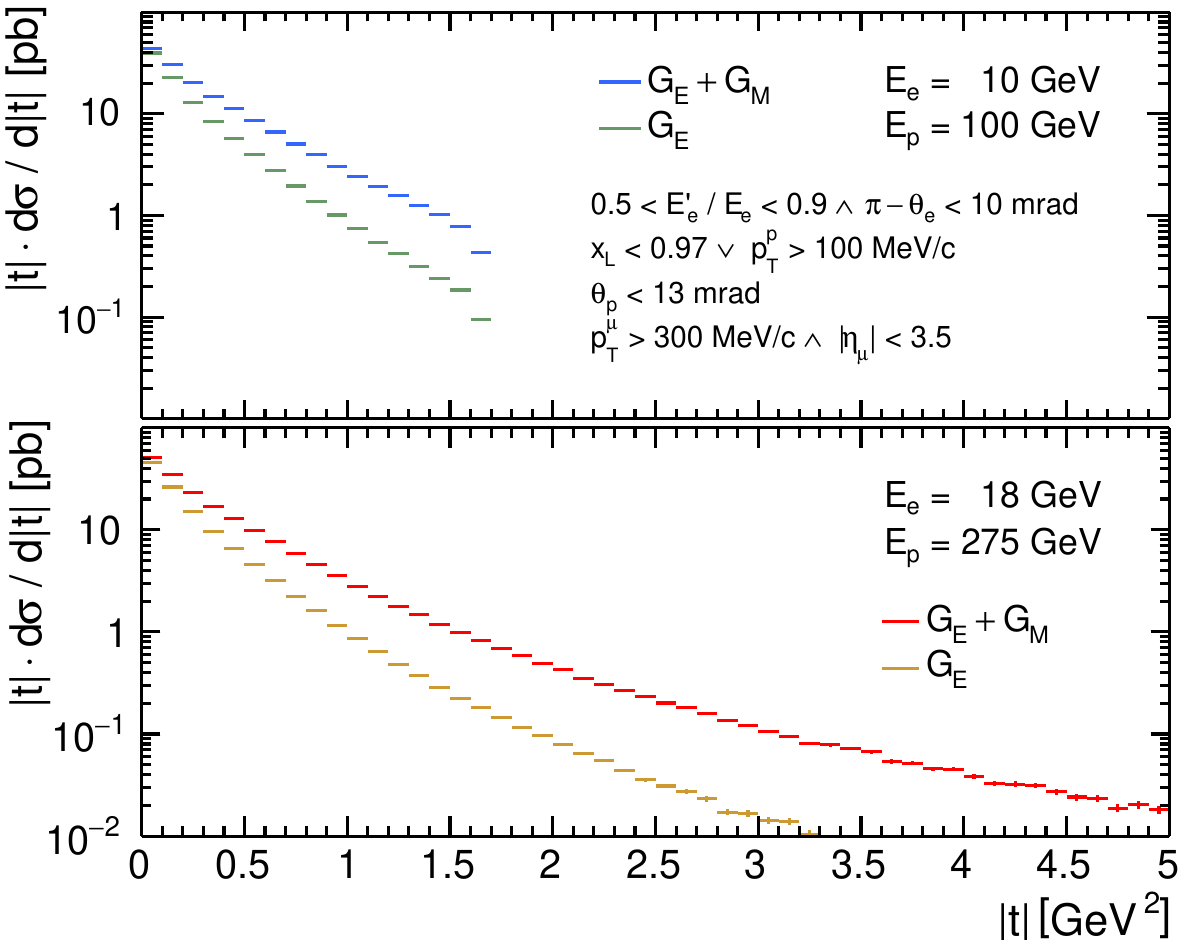} 
\caption{Differential cross-sections, 
$\lvert t\rvert\cdot \rm{d}\sigma/\rm{d}\lvert t\rvert$,
for accepted the events without the photon veto; for low (upper) and high 
(lower) energy beams, where the statistical errors correspond to the integrated 
luminosity of about 300~fb$^{-1}$. 
The plots compare distributions obtained using the full form-factor and 
those for which the magnetic form-factor was switched off.}  \label{fig5b}
\end{figure}

In Fig. \ref{fig5b}, the observed $\lvert t\rvert\cdot \rm{d}\sigma/\rm{d}\lvert t\rvert$ 
distributions are shown for electron-proton scattering, demonstrating the large $\lvert t\rvert$ reach also at low energy EIC running. At high energies a significant cross-section is expected even at $\lvert t\rvert\approx4~{\rm GeV}^2$. One should note that also here the requirement of electron detection could be lifted, leading to a large increase of the observed cross-section and to a lack of sensitivity to the bremsstrahlung overlays in the far-backward detectors. At such a high $\lvert t\rvert$, the $G_M$ contribution dominates but the $G_E$ one is still significant allowing their separation by combining data at different beam energies. Moreover, the high polarization of incident protons will enable construction of the azimuthal $p-\mu\mu$ correlations, for example, to enhance the separation power of the $G_E$ and $G_M$ contributions. 

As in the case of the proton charge radius, the EIC ion beams will allow us to extend 
such a studies by supplementing them with unique investigations of the nuclear effects. 
It is particularly exciting for the light ion beams for which a high polarization will 
be available \cite{CDR}.

\subsection{The exclusive pairs of tau leptons}\label{sec3.4}
Exclusive production of $\tau$ pairs via photon-photon fusion in the Ultra Peripheral Collisions (UPC) of heavy ions has recently become a vigorous research activity \cite{upc}. It was shown that such a two-photon pair production is particularly sensitive to the anomalous electromagnetic dipole moments of $\tau$ leptons. Here we evaluate the EIC scientific potential in this domain.

\begin{figure}[h]%
\centering
\includegraphics[width=0.6\textwidth]{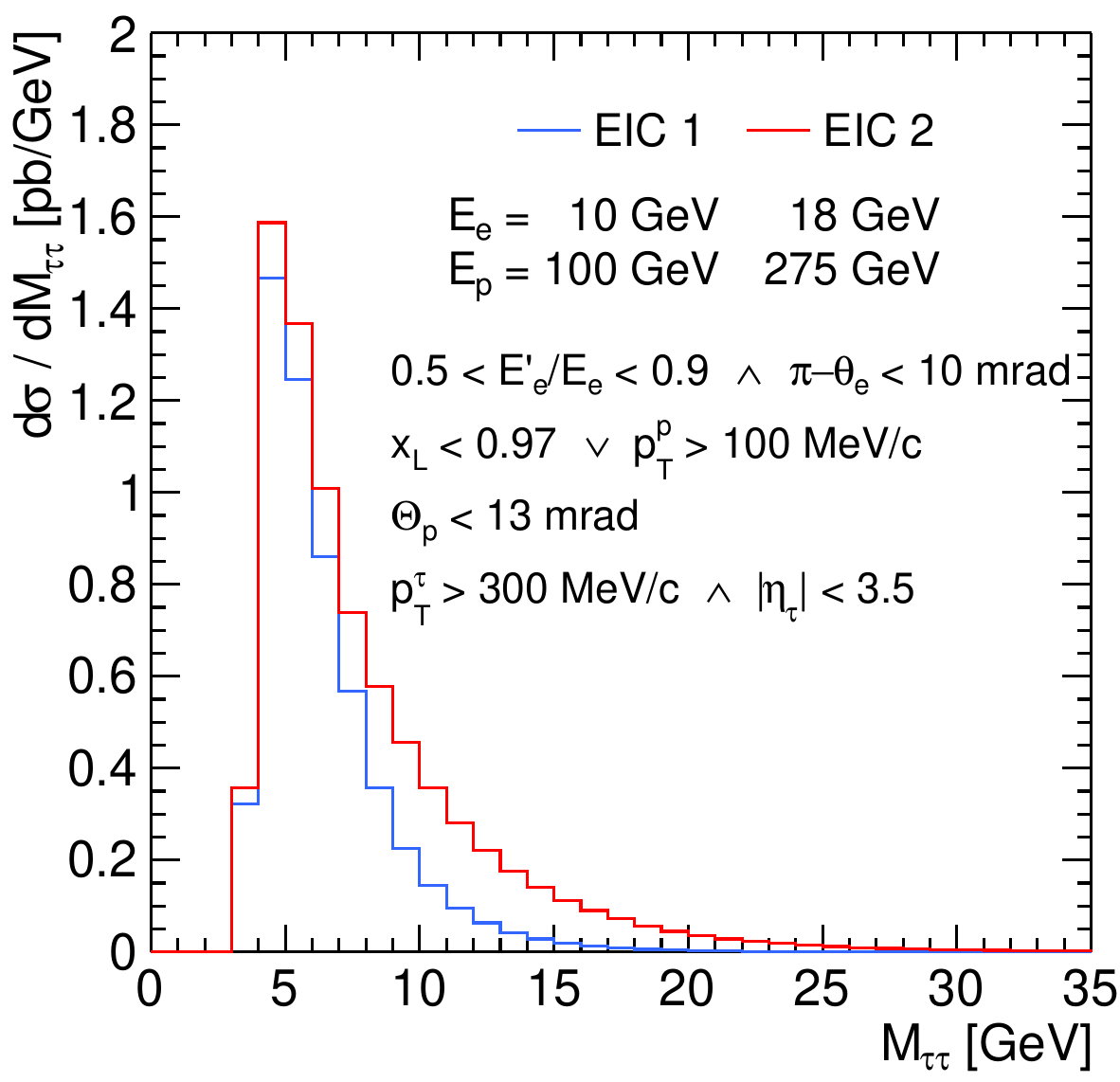}
\caption{The differential cross-section in the invariant mass $M_{\tau\tau}$ of selected events according to ``acceptance cuts" (see Sec. \ref{sec2}, without photon veto) for electron proton collisions at high (red histogram) and low (blue histogram) beam energies.
}  \label{fig6}
\end{figure}

In Fig. \ref{fig6} the differential cross-sections in the invariant mass $M_{\tau\tau}$ 
are shown for the $\tau^+\tau^-$ pairs of the accepted events (without imposing the 
photon veto). The corresponding total observed cross-sections are: 
$\sigma_\mathrm{obs}^{\tau\tau}(\mathrm{EIC}~1) = 5.5\; \mbox{pb}$ and 
$\sigma_\mathrm{obs}^{\tau\tau}(\mathrm{EIC}~2) = 7.8\; \mbox{pb}$. These cross-sections are large in spite of requiring the detection of the scattered electron and proton, to ensure the best reconstruction of the event kinematics. As a result, the expected $\tau^+\tau^-$ event samples at the EIC are very large -- about two orders of magnitude bigger than those expected to be collected at the High Luminosity Large Hadron Collider (HL-LHC)~\cite{szczurek}. One should note that in Fig. \ref{fig6} the detection of all $\tau$ leptons produced within ``geometrical acceptance'' of the EIC central detectors is assumed. In contrast to the UPC case, this should not be far from reality given the lack of trigger inefficiencies at the EIC, thanks to the data streaming, and the central detectors optimized for low and medium \pT particle reconstruction and identification. 

The detection of the scattered electron might not be necessary, in which case the event statistics will increase by an order of magnitude. Such experimental issues will be further studied, as well as the use of the proton beam polarization to amplify the  sensitivity to the the anomalous electromagnetic dipole moments of $\tau$ leptons at the EIC.
\section{Conclusions and outlook}\label{sec4}
The first exploratory survey of the exclusive lepton pair production at the EIC reveals excellent prospects for studying such processes. The combination of very high luminosities, very hermetic high resolution detectors, clean experimental environment, high beam polarisations and large variety of ion beams provides ideal conditions for further investigations. 

The exclusive lepton pairs at the EIC are a powerful tool offering profound insights into the structure of hadrons as well as unique studies of the $\tau$ lepton properties.
\clearpage

\bibliographystyle{unsrturl}
\bibliography{grape-bibliography}

\end{document}